\documentclass[]{emulateapj}
\usepackage{psfig}
\def\ls{\mathrel{\mathchoice {\vcenter{\offinterlineskip\halign{\hfil
$\displaystyle##$\hfil\cr<\cr\sim\cr}}}
{\vcenter{\offinterlineskip\halign{\hfil$\textstyle##$\hfil\cr
<\cr\sim\cr}}}
{\vcenter{\offinterlineskip\halign{\hfil$\scriptstyle##$\hfil\cr
<\cr\sim\cr}}}
{\vcenter{\offinterlineskip\halign{\hfil$\scriptscriptstyle##$\hfil\cr
<\cr\sim\cr}}}}}

\def\ifm#1{\relax\ifmmode#1\else$\mathsurround=0pt #1$\fi}

\def\ltsima{$\; \buildrel < \over \sim \;$}
\def\lsim{\lower.5ex\hbox{\ltsima}}
\def\gtsima{$\; \buildrel > \over \sim \;$}
\def\gsim{\lower.5ex\hbox{\gtsima}}

\newenvironment{inlinefigure}{
\def\@captype{figure}
\noindent\begin{minipage}{0.999\linewidth}\begin{center}}
{\end{center}\end{minipage}\smallskip}

\lefthead{MOUSTAKAS \& IMMLER}
\righthead{}

\shorttitle{X-ray constraints on ionizing photons at Z$\sim$6}
\shortauthors{Moustakas \& Immler}

\begin{document}

\submitted{Submitted to the ApJL}

\title{X-ray constraints on ionizing photons from accreting black
holes at ${\rm Z}\sim6$\altaffilmark{1}}

\author{Leonidas A. Moustakas\altaffilmark{2}, 
Stefan Immler\altaffilmark{3}}
\altaffiltext{1}{Based on observations taken with the NASA/ESA Hubble
  Space Telescope, which is operated by the Association of
  Universities for Research in Astronomy, Inc.\ (AURA) under NASA
  contract NAS\,5--26555}
\altaffiltext{2}{Space Telescope Science Institute, 3700 San Martin
  Drive, Baltimore, MD 21218, {\tt leonidas@stsci.edu}}
\altaffiltext{3}{Laboratory for High Energy Astrophysics, Universities 
  Space Research Association, Code 662, NASA Goddard Space Flight
  Center, Greenbelt, MD 20771, {\tt immler@milkyway.gsfc.nasa.gov}}  

\begin{abstract}
Using an X-ray stacking procedure, we provide a robust upper limit to
the X-ray luminosity per object of a set of 54 $z\approx5.8$ galaxy
candidates in the \emph{Hubble} Ultra Deep Field (HUDF), which is
within the 1\,Ms-exposure \emph{Chandra} Deep Field-South (CDF-S).
With an effective total exposure of $44$\,Ms for the stack, the
$3\,\sigma$ flux-density limit of $2.1\times10^{-17}\,{\rm erg}\,{\rm
cm}^{-2}\,{\rm s}^{-1}$ (soft-band) gives a $3\,\sigma$ upper-limit
luminosity of $L_{\rm X} = 8\times10^{42}$\,erg\,s$^{-1}$ per object
at a rest-frame hard energy range of $\sim3$--$14$\,keV at
$z\approx5.8$ for a photon index of $\Gamma=2$.  For an active
accreting black hole (or ``mini-quasar'') emitting at the Eddington
luminosity ($\eta_{\rm Edd}=1$), and the Sazonov~et\,al.~average-QSO
spectral energy distribution, we calculate an upper limit on the black
hole mass, $M_{\rm bh}< 3\times10^{6}\,\eta_{\rm Edd}^{-1}\,M_{\odot}$
($3\,\sigma$).  The X-ray limit further implies an upper limit on the
rate density of UV ionizing photons from accreting black holes at that
redshift, $\dot{n}_{\rm ioniz}<2\times10^{51}\,{\rm s}^{-1}\,{\rm
Mpc}^{-3}$ ($3\,\sigma$), which is less than 1/10 of the number needed
to ionize the universe.  Because the constraint is anchored in the
rest-frame hard X-ray regime, a steeper spectrum for ``mini-quasars''
would imply relatively \emph{fewer} ionizing UV photons.  Unless there
are large populations of active black holes around this mass that are
\emph{unassociated} with luminous galaxies, mini-quasars do not appear
to contribute significantly to the budget of ionizing photons at
$z\approx6$.
\end{abstract}


\section{Introduction}\label{sec:intro}

The identification of the epoch or epochs of reionization is not a
solved problem, though there is tension between the high redshift
($z_{\rm reion}\sim10$--$17$) of reionization implied by the WMAP
results \citep{spergel:03, kogut:03} and the lower redshift ($z_{\rm
reion}\approx6$) suggested by the detection of the Gunn-Peterson trough
in $z\approx6$ luminous QSOs by SDSS \citep{becker:01, djorgovski:01}.
These observations suggest that the transition from an opaque to a
fully reionized universe must have at least \emph{ended} by
$z\approx6$.  One of the challenges faced today is to identify the
sources of the ionizing photons \citep[e.g.][and references
therein]{loeb:01, stiavelli:04a}, which are still unaccounted for from
observed stellar and accreting black-hole (``active'') populations.

The Advanced Camera for Surveys (ACS; \citealt{benitez:04}) on the
\emph{Hubble Space Telescope} (\emph{HST}) has made routine the
discovery of candidate galaxies at redshift $z\approx6$
\citep{bunker:03, bouwens:03, stanway:03, stanway:04, dickinson:04,
giavalisco:04b}.  These objects are identified by an extension of the
``Lyman Break'' technique \citep{steidel:92, steidel:96}, whereby the
opacity from neutral hydrogen in the intergalactic medium (or the
Lyman-$\alpha$ forest) suppresses the flux at wavelengths shortward of
the Lyman break \citep{madau:96}, or shortward of the Lyman-$\alpha$
emission \citep{dickinson:04}).  The \emph{HST}/ACS $i_{775}$ and
$z_{850}$ filters bracket Lyman-$\alpha$ ($\lambda_{\rm
rest}=1216$\,\AA) around $z\approx5.8$, so searches for
``$i$-dropouts'' will isolate candidate galaxies around that redshift
\citep{stanway:03, dickinson:04}, as well as possible ``interlopers''
in the guise of cool stellar dwarfs or faint evolved galaxies at
$z\approx1$--$2$.  Some of these high-redshift candidate galaxies have
been confirmed spectroscopically from the ground, and from the G800L
grism of ACS \citep{pirzkal:04, malhotra:04}.

Using data from the ACS-based \emph{Hubble} Ultra Deep Field (HUDF;
\citealt{beckwith:04}), \citet{bunker:04} identified 54
spatially-resolved $i$-dropout candidates to $z_{850}\approx28.5$.  It
is still an open question whether some combination of these galaxies,
possibly with the contribution of as yet unveiled populations, have
enough energetic photons to ionize (or keep ionized) the universe at
this epoch \citep{bunker:04, stiavelli:04b}.

The need to account for the large number of required ultraviolet
photons for ionization \citep*{madau:99} has given rise to a broad
range of proposed sources, ranging from known candidates (stellar and
active sources) to exotic ones \citep[e.g.~decaying heavy
neutrinos,][]{hansen:04}.  The stellar sources can be divided broadly 
into two categories.  The first-generation ``metal-free'' stars
\citep{abel:02}, which may arise in $M_{\rm dm}\gsim10^{5}\,M_{\odot}$
dark matter halos, are expected to have masses of
$M_{*}\gsim60M_{\odot}$\footnote{Such stars with masses in the range
$M_{*}\sim60$--$140M_{\odot}$, and $M_{*}>260M_{\odot}$ are predicted
to collapse directly to a black hole \citep{heger:03}, with
$M_{*}\sim140$--$260$\,$M_{\odot}$ stars annihilating themselves
through pair-creation processes.  Such collapsed stars may be the
seeds of the ``mini-quasar'' black holes discussed below.}.  These
stars, though rare, may produce enough photons to contribute most or
all of the UV photons needed to ionize the universe even at very early
times ($z\sim20$; e.g.~\citealt{rss:03a}).  If such objects are to be
found at $z\sim6$, it is unlikely that significant amounts of X-ray
emission would be associated with them.  Population~II stars, which
may also contribute to the ionizing radiation budget at high redshift,
could have associated X-ray emission from X-ray binaries that result
from different modes of star formation \citep{grimm:03}.  Though such
relatively locally determined relations may not hold over all time,
they can be used to explore relevant limits to star formation rates at
high redshift.

It is plausible that many of the first collapsed objects produced
black holes of some mass, that could then be sources of energy.  This
possibility is potentially attractive as a source of energetic
photons, particularly if there are mechanisms for self-regulation of
their growth \citep{wyithe:03b}.  The mass-increase $e$-folding time
of black holes emitting at their Eddington luminosities, the Salpeter
timescale, is $T_{\rm Salp} \approx
1\times10^{8}\,(\epsilon/0.30)^{-1}\,{\rm yr}$, where $\epsilon$ is
the energy conversion efficiency \citep{elvis:02, gammie:04}.  The
$T_{\rm Salp}$ is independent of the black hole mass.  If the active
lifetime of an accreting black hole at $z\sim6$ is several times
$T_{\rm Salp}$, so that the probability of any one host galaxy being
observed while ``active'' is high, it is possible that the majority of
galaxies hosting black holes could be observed during an active stage.

In this paper we use one plausible set of dropout-selected galaxies at
$z\approx5.8$, described in \S\,\ref{sec:hudf}.  None of these sources
are individually detected in the deep \emph{Chandra} observations in
that field.  We therefore construct an extremely deep stack of the
full sample, as set out in \S\,\ref{sec:stack}, and compute an upper
limit to the average X-ray luminosity for each galaxy.  This number is
cast against proposed sources of ionizing photons at that redshift in
\S\,\ref{sec:ioniz}, and the possible utility of these limits are 
discussed in \S\,\ref{sec:disc}.  We assume the concordance cosmology
parameters of a flat universe with
$\Omega_{\Lambda}=1-\Omega_{m}=0.7$, and
$H=70\,h_{70}$\,km\,s$^{-1}$\,Mpc$^{-1}$.

\section{The HUDF and high-redshift candidates}\label{sec:hudf}

The HUDF\footnote{http://www.stsci.edu/hst/udf/} is a
11.3\,arcmin$^2$ ACS-based survey of a single extremely deep field
observed within the
GOODS-South\footnote{http://www.stsci.edu/science/goods/} field
\citep{giavalisco:04a}.  The observations have been made in the same
filters as the GOODS project, $B_{435}$, $V_{606}$, $i_{775}$ and
$z_{850}$, and reach $10\sigma$ depths of $z_{850}\approx28.5$ (AB).
This is also the 1\,Ms \emph{Chandra} Deep Field-South (CDF-S) field,
which reaches on-axis flux density limits of
$5.5\times10^{-17}$\,erg\,s$^{-1}$\,cm$^{-2}$ in the soft-band
(0.5--2\,keV) and $4.5\times10^{-16}$\,erg\,s$^{-1}$\,cm$^{-2}$ in the
hard-band (2--8\,keV; \citealt{giac:02}).

By virtue of the dropout selection with the $i_{775}$ and $z_{850}$
bandpasses, the redshift distribution/selection function of the
high-redshift galaxies is quite narrow, centered around $z\approx5.8$.
The candidate list in \citet{bunker:04} is a lower limit on the total
number visible in the HUDF, because of the very conservative
signal-to-noise requirements applied in the detection band.  At the
same time, by the strong color criterion, there should be few
foreground ($z\approx2$) galaxies.  Therefore we adopt the tabulated
list of candidates directly from \citet{bunker:04}, several of which
have already been spectroscopically confirmed to be at the claimed
redshift range.

\begin{inlinefigure}
\begin{center}
\resizebox{\textwidth}{!}{\includegraphics{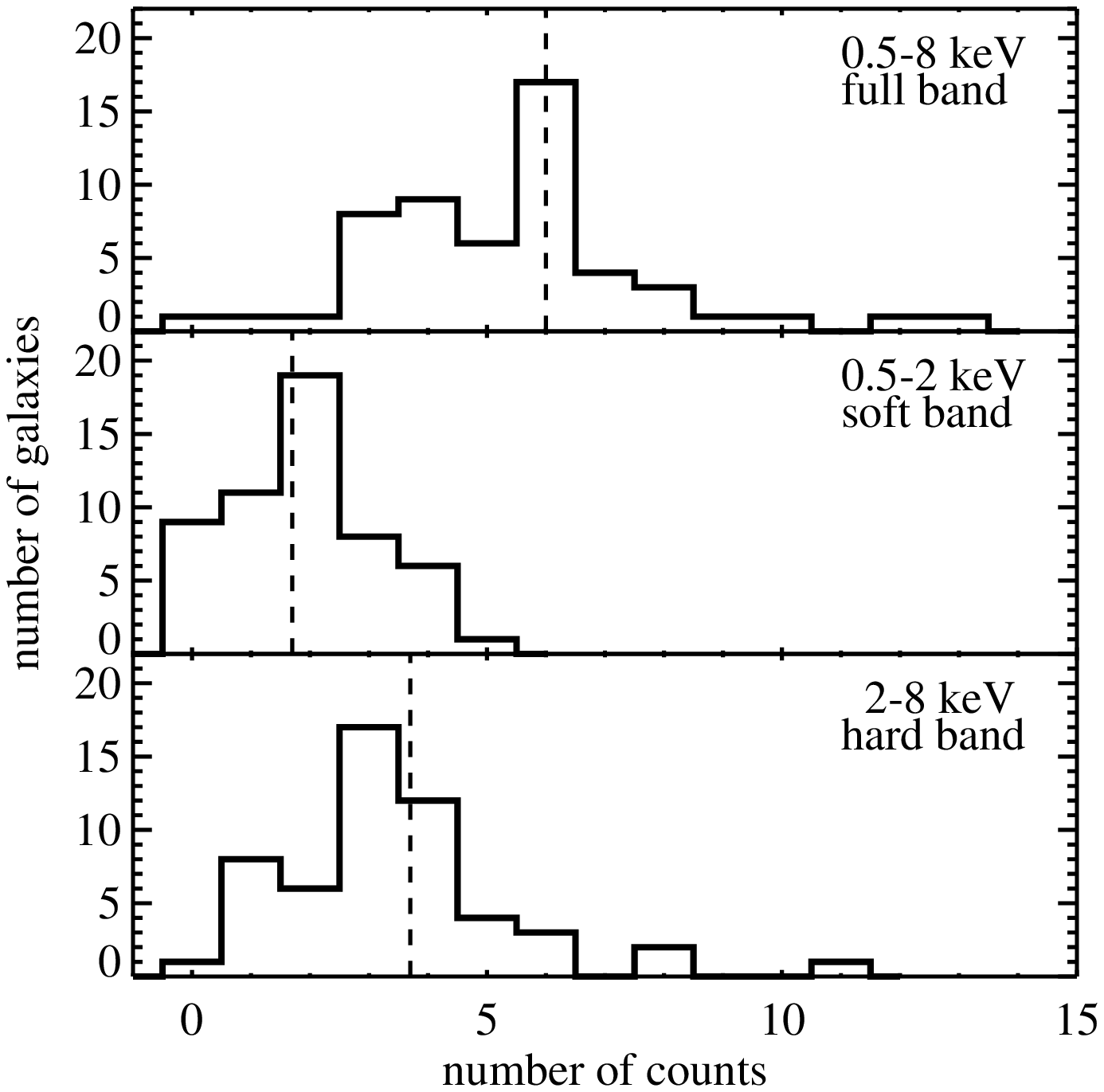}}
\figcaption{The distribution of X-ray counts for all candidate
$z\approx5.8$ galaxies.  The vertical dashed lines mark the mean
background count levels in each band, as determined from Monte Carlo
simulations with 10,000 trials.  \label{fig:counts}}
\end{center}
\end{inlinefigure}
 
\begin{center}
\begin{table}
\begin{center}
\caption{Results from the X-Ray Stacking \label{tab:stack}}
\begin{tabular}{l ccc cc}
 \multicolumn{1}{c}{Band} &
 \multicolumn{1}{c}{Counts} &
 \multicolumn{1}{c}{$T_{\rm eff}$} & 
 \multicolumn{1}{c}{Rate} &
 \multicolumn{1}{c}{$f_{\rm x}$ ($3\,\sigma$)} &
 \multicolumn{1}{c}{$L_{\rm x}$ ($3\,\sigma$)} \\
 \multicolumn{1}{c}{} &
 \multicolumn{1}{c}{} &
 \multicolumn{1}{c}{(Ms)} &
 \multicolumn{1}{c}{($10^{-6}$)} &
 \multicolumn{1}{c}{($10^{-17}$)} &
 \multicolumn{1}{c}{($10^{43}$)} \\
 \multicolumn{1}{c}{(1)} & (2) & (3) & (4) & (5) & (6) \\
\hline
\hline
0.5--8           & $319.1\pm19.8$ & $44.3$ & $7.20\pm0.45$ & \phantom{0}$8.0\pm0.5$ &  $3.16$ \\
0.5--2           & $114.3\pm11.9$ & $44.1$ & $2.59\pm0.27$ & \phantom{0}$2.1\pm0.2$ &  $0.83$ \\
\phantom{0.}2--8 & $204.7\pm15.7$ & $43.5$ & $4.70\pm0.36$ &           $11.7\pm0.9$ &  $4.68$ \\
\hline
\end{tabular}
\end{center}
\tablenotetext{}{(1)~Observed-frame energy band $E_{\rm
  obs}$ in keV; (2)~Total aperture counts; fractional counts are due
  to fractional detection cells within the apertures; (3)~Effective
  exposure time; (4)~Exposure-corrected count rate in $10^{-6}~{\rm
  s}^{-1}$; (5)~X-ray fluxes in ${\rm erg~cm}^{-2}~{\rm s}^{-1}$,
  $\Gamma=2$; (6)~Rest-frame X-ray luminosities at energy bands
  $(1+z)\times\,E_{\rm obs}$ for $z=5.8$, in ${\rm erg~s}^{-1}$, for
  $\Gamma=2$.}
\end{table}
\end{center}

\section{The X-ray Stack Procedure and Results}\label{sec:stack}

Stacking the X-ray images of individually undetected objects has
offered exceptional insights into the nature of star formation and
activity in otherwise X-ray-invisible extremely distant objects
\citep[e.g.][]{davo:02b, davo:02a, nandra:02, reddy:04}.  In the
present work, we target the positions of all the sources described in
the previous section.  A search radius of $1.5$ times the error in the
X-ray position was applied to match each of the optical positions with
the X-ray source catalogs by \citet{giac:02} and \citet{bauer:04}.
For each object position, exposure-corrected counts were extracted
within the 90\% encircled-energy (EE) radii in the full (0.5--8\,keV),
soft (0.5--2\,keV), and hard (2--8\,keV) energy bands.  Also at each
position, in order to quantify the significance (or lack thereof) of
the counts measurement, we perform Monte Carlo (MC) simulations with
10,000 trials using source-free circular regions of $20$\,arcsec
radius around each galaxy.  The net aperture counts, then, are
obtained by subtracting the mean value of the (near-Gaussian) MC
distribution from the number of average aperture counts. Details of
the stacking procedure are discussed in \citet{immler:04}.

Using the above apertures to count the number of X-ray photons at each
galaxy position, we detect 319~counts in total, distributed as shown
in Fig.~\ref{fig:counts} and tabulated in Table~\ref{tab:stack}.  The
exposure-corrected count rate in the full band is
$\sim2.6\times10^{-6}$\,s$^{-1}$ in the soft band, with a total
effective exposure time of 44.1\,Ms.  There is no formal X-ray
detection in the stacked data.  Assuming $z=5.8$, a power law spectral
energy distribution (SED) with a photon index of $\Gamma=2$, and a
foreground absorbing column density of $N_{\rm HI} =
8.0\times10^{19}\,{\rm cm}^{-2}$ \citep{stark:92}, we find a
band-corrected $3\,\sigma$ limit on the luminosity \emph{per galaxy} of
$L_{\rm X}<8.3\times10^{42}\,{\rm erg}\,{\rm s}^{-1}$ ($3\,\sigma$) in
the rest-frame energy band $(1+z)\times(0.5$--$2\,{\rm keV}) \approx
3$--$14\,{\rm keV}$.  The conversion between counts and flux is not
very sensitive to the assumed spectral properties --- a photon index
of $\Gamma=1.7$ instead of $\Gamma=2$ gives a flux which is $\ls5\%$
lower.  As the $k$-correction for a power law with photon index
$\Gamma$ is $k(z)=(1+z)^{\Gamma-2}$, the X-ray luminosity constraints
can correspondingly be lower by $\sim40$\%\ for $\Gamma=1.7$.

\begin{inlinefigure}
\begin{center}
\resizebox{\textwidth}{!}{\includegraphics{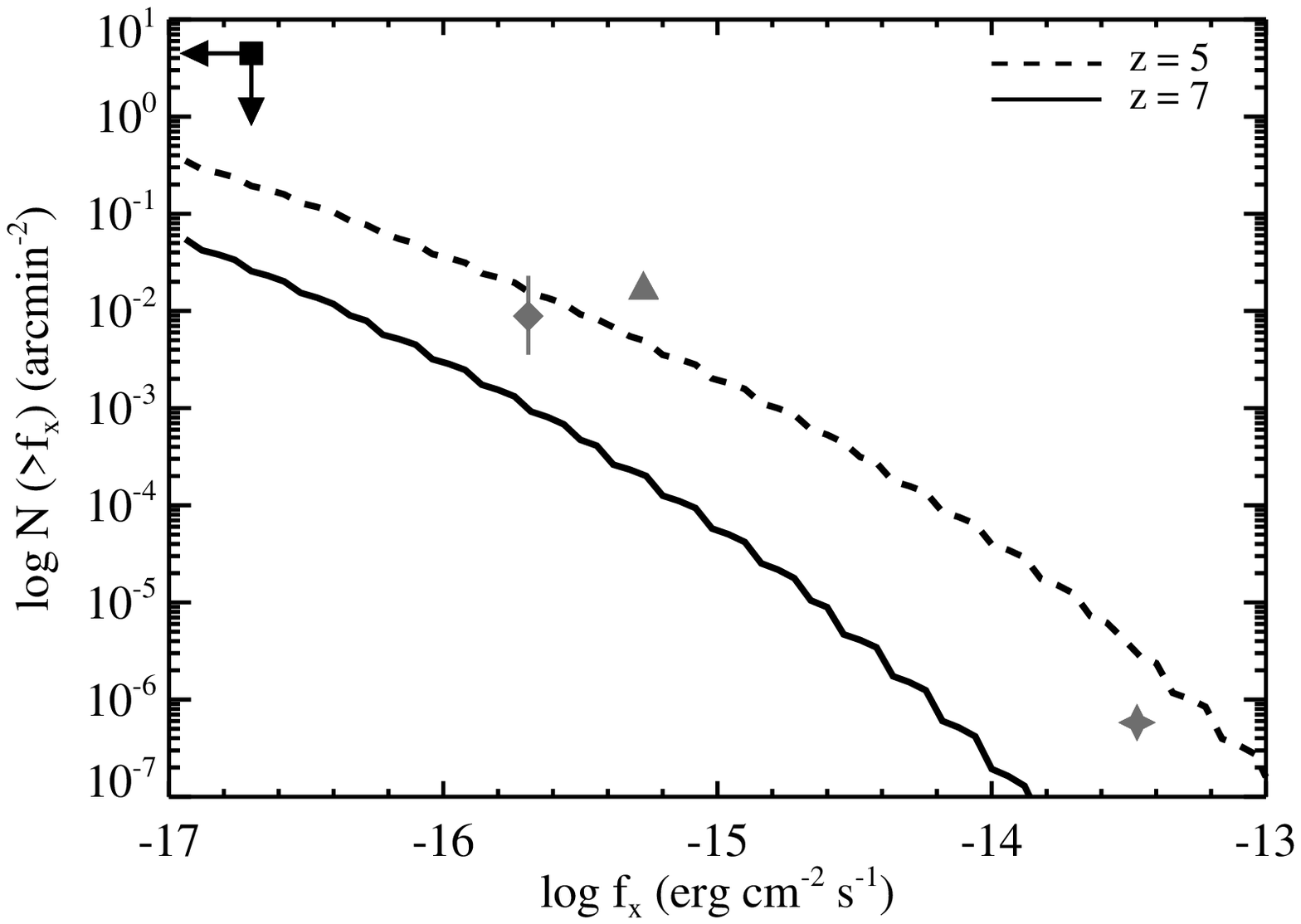}}
\figcaption{The cumulative surface density of X-ray emitting active
objects versus flux limit.  The model lines are drawn for quasar
number counts at redshifts of $z=5$ (dashed line) and $z=7$ (solid
line) from \citet{wyithe:03b}.  The limits determined from this study
for $z\approx5.8$ are shown (black box).  The counts from
\citet{barger:03} at $z>5$ (diamond), \citet{cristiani:04} at
$3.5\gsim\,z\gsim5.2$ (triangle), and \citet{fan:04} at $z\gsim5.7$
(star) are shown for comparison.
  \label{fig:numdens}} 
\end{center}
\end{inlinefigure}
 
\section{High-redshift sources of ionizing photons}\label{sec:ioniz}

There is still no consensus on the source or sources of ionizing
photons at redshifts $z\sim6$ and beyond \citep[see][for a review of
candidate populations]{loeb:01}, though it is possible that a complex
reionization history \citep{cen:03, hui:03} could have been driven by
different combinations of ionizing sources at different times
\citep[e.g.][]{rss:03a, sokasian:03, sokasian:04, madau:04}.  The
highest redshift where there are presently extensive data is
$z\approx6$.  The success of identifying how the universe stays
ionized at that redshift is mixed \citep{yan:04, bunker:04,
stiavelli:04b}.  In this section, we consider the implications of our
measured limits.

\subsection{Star formation}

Without making any corrections for obscuration, the \citet{kenn:98}
relation ${\rm SFR}_{\rm UV} (M_{\odot}\,{\rm yr}^{-1}) =
1.4\times10^{-28} L_{\nu} ({\rm erg}\,{\rm s}^{-1}\,{\rm Hz}^{-1})$
(for a Salpeter IMF) gives a typical ${\rm SFR}_{\rm
UV}\approx10\,M_{\odot}\,{\rm yr}^{-1}$ for each galaxy, where we have
used the $z_{850}$ photometry ($\lambda_{\rm rest}\approx1400$\,\AA).
At low redshift, there is a well-explored association between star
formation rates and X-ray luminosity, driven by the aggregate effect
of low- and high-mass X-ray binaries (LMXBs and HMXBs, respectively),
which are formed at different stages of stellar evolution.  At star
formation rates above several~M$_{\odot}$\,yr$^{-1}$, promptly-formed
HMXBs dominate, and \citet{grimm:03} find ${\rm SFR}_{\rm
X}\,(M_{\odot}\,{\rm yr}^{-1}) = L_{\rm 2-8
keV}\,/\,(6.7\times10^{39}\,{\rm erg}\,{\rm s}^{-1})$.  Using the
rest-frame hard X-ray limit, we calculate ${\rm SFR}_{\rm
X}<1100\,M_{\odot}\,{\rm yr}^{-1}$ ($3\,\sigma$).  This limit is
therefore not useful compared to the measurements from the rest-UV
\citep{bunker:04, stiavelli:04b}, and we do not discuss it further.

\subsection{Accreting black holes}

There is theoretical motivation for black hole accretion-powered
``mini-quasars'' at high redshift \citep[e.g.][]{madau:04}.  These
could begin forming at very early times ($z\gsim20$) from the collapse
of first-generation massive stars, and grow substantially in mass by
$z\approx6$.  Whereas the comoving space density of
\emph{supermassive} ($M_{\rm bh}\gsim10^{9} M_{\odot}$) black holes at
this redshift is exceedingly small \citep{fan:03}, there are only
inferences on the numbers of lower-mass black holes
\citep[e.g.][]{wyithe:03b, bromley:04}.

With this noted, however, we are encouraged to compute what black hole
masses the X-ray limits in Table~\ref{tab:stack} imply, assuming that
these putative black holes are actually active.  For the purposes of
this calculation, we scale the bolometric luminosity to the Eddington
luminosity, as $L_{\rm bol}=\eta_{\rm Edd}L_{\rm Edd}$.  Since our
observations are in a restricted energy band, we use a fiducial SED
template \citep{sazonov:04} to calculate the fraction of the total
energy in the relevant band, $f_{\rm b}$ (the inverse of which is the
usual bolometric correction factor).  Then the luminosity in band $b$
is $L_{\rm X,b}=f_{\rm b}\,\eta_{\rm Edd}\,L_{\rm Edd}$, or
\begin{eqnarray}
L_{\rm X,b} & = & 2.52\times10^{42}\,(f_{\rm b}/0.02)\,(\eta_{\rm
Edd}/1.0)\times \nonumber\\
            &   & (M_{\rm bh}/10^{6}\,M_{\odot})\,{\rm erg}\,{\rm s}^{-1},
\end{eqnarray}
where $f_{\rm b}\approx0.02$ is calculated for the observed soft-band
at $z\approx5.8$ directly from the template.  Assuming the best case
of $\eta_{\rm Edd}\approx1.0$ \citep[e.g.][]{floyd:04}, we find an
upper limit on the average mass of \emph{active} black holes in each
galaxy of 
\begin{equation}
M_{\rm bh} < 3.3\times10^{6}\,(f_{\rm b}/0.02)^{-1}\,(\eta_{\rm
  Edd}/1.0)^{-1}\,M_{\odot}~({\rm 3\,\sigma})\label{eq:mbh}. 
\end{equation}

Integrating the \citet{sazonov:04} template between $1$--$4$\,ryd
(between the H\,{\scriptsize I} and He\,{\scriptsize II} edges), we
find an energy fraction of $f_{\rm UV}\approx0.056$.  The production
rate limit of H\,{\scriptsize I}-ionizing photons per galaxy is then
$\dot{N}_{\rm ioniz}<L_{\rm X}\,(f_{\rm UV}/f_{\rm
b})/\langle h\nu\rangle\approx4.0\times10^{53}\,{\rm s}^{-1}.$
Considering the ensemble of galaxies and the approximate relevant
volume, we estimate the ionizing photon rate density
\begin{equation}
\dot{n}_{\rm ioniz}<2.0\times10^{51}\,{\rm s}^{-1}\,{\rm Mpc}^{-3}
(3\,\sigma) \label{eq:nion}. 
\end{equation}

\section{Discussion}\label{sec:disc}

If remnants from $z\approx20$ $M_{\rm st}\sim100\,M_{\odot}$
population~III stars form the seeds of early black holes, which then
grow through radiative accretion until $z\approx6$, they would have
masses $M_{\rm bh}\approx10^{5-6}\,M_{\odot}$, even for a large
$\epsilon\approx0.3$ and before considering merging.  However, at the
limit of Eq.~\ref{eq:mbh}, the density of (active) black hole mass
inferred is $\rho_{\rm bh,6}<1.5\times10^{4}\,M_{\odot}\,{\rm
Mpc}^{-3}$, which is much smaller than the present-day density of all
black holes, $\rho_{\rm bh,0}=2.9\times10^{5}\,M_{\odot}\,{\rm
Mpc}^{-3}$ \citep{yu:02}.  Therefore, the present limit is plausible,
while allowing great latitude for black holes to continue to grow (and
shine).

For full (hydrogen) ionization, the ionization rate
(Eq.~\ref{eq:nion}) must exceed the recombination rate, $\dot{n}_{\rm
recomb}=C\,\alpha_{\rm B}\,n_{\rm H}(z)^2$, where $C=\langle
n_e^2\rangle/\langle n_e\rangle^2$ is the clumpiness of the ionized
gas, $\alpha_{\rm B}\approx2.6\times10^{-13}\,{\rm cm}^3\,{\rm
s}^{-1}$ is the (case B) recombination coefficient for a temperature
$T\approx10^4$\,K, and $n_{\rm H}(z)$ is the particle density of
hydrogen.  Then,
\begin{eqnarray}
\dot{n}_{\rm recomb} & = & C\,\alpha_{\rm B}\,[(1-Y_{\rm He})\,n_{\rm
    b,0}\,(1+z)^3]^2  \nonumber \\
 & = & 3.2\times10^{52}\,(C/1.0)\,{\rm s}^{-1}\,{\rm Mpc}^{-3},
\end{eqnarray}
where $n_{\rm b,0}=2.7\times10^{-7}\,{\rm cm}^{-3}$ \citep{spergel:03}
is the present-day density of baryons, and $Y_{\rm He}\approx0.24$ is
the primordial He fraction.  Estimates and simulations suggest that
the clumpiness is $C\approx1$--$30$ \citep{madau:99, cen:03}.
Therefore, even at the upper limit of Eq.~\ref{eq:nion}, active black
holes associated with luminous galaxies cannot be responsible for but
a small fraction of the total ionizing flux at $z\approx6$.  Most
changes to the calculations would be in the sense of \emph{decreasing}
the $\dot{n}_{\rm ioniz}/\dot{n}_{\rm recomb}$ ratio.  Cosmic variance
\citep{barkana:03, rss:04b} could affect these results by a factor of
two.  The surface density of the $z\approx6$ galaxies used here is,
however, consistent with results from independent fields
\citep{bouwens:03}.

It is worth noting that if the SED of an accreting $M_{\rm
bh}\lsim10^{6}\,M_{\odot}$ black hole is \emph{harder} than the
\citet{sazonov:04} template, as has been argued e.g.~by
\citet{madau:04}, the X-ray limit derived above would imply
\emph{fewer} ionizing UV photons.  Such sources may still be playing
an important role with partial ionization (especially if they are not
explicitly associated with luminous galaxies), or may be instrumental
in achieving the early reionization suggested by WMAP
\citep[e.g.][]{rss:03b, madau:04}.

Our results are consistent with \citet{dijkstra:04}, who use the
observed soft X-ray background to infer the maximum contribution to
the rest-frame hard X-ray background at $z>6$, and based on strong
observational limits, rule out a significant contribution to the
ionizing flux by active accreting black holes.

Based on a self-regulated prescription for the growth of black holes
in a $\Lambda$CDM framework, \citet{wyithe:03b} motivate the local
\citet{magorrian:98} relation, and make predictions for massive black
holes as a function of redshift.  Their predicted number counts of
active black holes at high redshift are shown in
Figure~\ref{fig:numdens}, with the data from this work. This shows
that useful constraints on such $\Lambda$CDM models will be possible
with only a few times larger survey areas, even at sensitivities
comparable to the present CDF-S.  The \citet{wyithe:03b} black hole
mass function (see their fig.~5) predicts a cumulative black hole
number density of $n_{\rm bh}\approx9\times10^{-4}\,{\rm Mpc}^{-3}$
down to $M_{\rm bh}\approx3\times10^{6}\,M_{\odot}$, which is about
five times smaller than the upper limit from our sample.  This
suggests that, distinct from the census of ionizing photons from such
sources, larger deep \emph{Chandra} surveys in fields with large
high-redshift galaxy samples may achieve stronger constraints on the
models.  Future X-ray missions, such as \emph{Constellation-X} and
\emph{XEUS} will easily reach the relevant sensitivities -- if any
black holes driving ``mini-quasars'' at $z\approx6$ are there to be
found at all.

\acknowledgments 

We are grateful to R. Somerville, D. Stern, M. Livio, T. Abel, and
C. Vignali for discussions and comments, and to S. Wyithe and A. Loeb
for sharing their model predictions.  L.A.M. acknowledges support by
NASA through contract number 1224666 issued by the Jet Propulsion
Laboratory, California Institute of Technology under NASA contract
1407.

\bibliographystyle{apj}
\bibliography{udfx}

\begin{thebibliography}{59}
\expandafter\ifx\csname natexlab\endcsname\relax\def\natexlab#1{#1}\fi

\bibitem[{{Abel} {et~al.}(2002){Abel}, {Bryan}, \& {Norman}}]{abel:02}
{Abel}, T., {Bryan}, G.~L., \& {Norman}, M.~L. 2002, Science, 295, 93

\bibitem[{{Alexander} {et~al.}(2002{\natexlab{a}}){Alexander}, {Aussel},
  {Bauer}, {Brandt}, {Hornschemeier}, {Vignali}, {Garmire}, \&
  {Schneider}}]{davo:02b}
{Alexander}, D.~M., {Aussel}, H., {Bauer}, F.~E., {Brandt}, W.~N.,
  {Hornschemeier}, A.~E., {Vignali}, C., {Garmire}, G.~P., \& {Schneider},
  D.~P. 2002{\natexlab{a}}, ApJL, 568, L85

\bibitem[{{Alexander} {et~al.}(2002{\natexlab{b}}){Alexander}, {Vignali},
  {Bauer}, {Brandt}, {Hornschemeier}, {Garmire}, \& {Schneider}}]{davo:02a}
{Alexander}, D.~M., {Vignali}, C., {Bauer}, F.~E., {Brandt}, W.~N.,
  {Hornschemeier}, A.~E., {Garmire}, G.~P., \& {Schneider}, D.~P.
  2002{\natexlab{b}}, AJ, 123, 1149

\bibitem[{{Barger} {et~al.}(2003){Barger}, {Cowie}, {Capak}, {Alexander},
  {Bauer}, {Brandt}, {Garmire}, \& {Hornschemeier}}]{barger:03}
{Barger}, A.~J., {Cowie}, L.~L., {Capak}, P., {Alexander}, D.~M., {Bauer},
  F.~E., {Brandt}, W.~N., {Garmire}, G.~P., \& {Hornschemeier}, A.~E. 2003,
  ApJL, 584, L61

\bibitem[{{Barkana} \& {Loeb}(2003)}]{barkana:03}
{Barkana}, R. \& {Loeb}, A. 2003, astro-ph/0310338

\bibitem[{{Bauer} {et~al.}(2004)}]{bauer:04}
{Bauer}, F.~E. {et~al.} 2004, ApJS, in prep.

\bibitem[{{Becker} {et~al.}(2001)}]{becker:01}
{Becker}, R.~H. {et~al.} 2001, AJ, 122, 2850

\bibitem[{{Beckwith} {et~al.}(2004)}]{beckwith:04}
{Beckwith}, S. {et~al.} 2004, ApJ, in prep.

\bibitem[{{Ben{\'{\i}}tez} {et~al.}(2004)}]{benitez:04}
{Ben{\'{\i}}tez}, N. {et~al.} 2004, ApJS, 150, 1

\bibitem[{{Bouwens} {et~al.}(2003)}]{bouwens:03}
{Bouwens}, R.~J. {et~al.} 2003, ApJ, 595, 589

\bibitem[{{Bromley} {et~al.}(2004){Bromley}, {Somerville}, \&
  {Fabian}}]{bromley:04}
{Bromley}, J.~M., {Somerville}, R.~S., \& {Fabian}, A.~C. 2004, MNRAS, 350, 456

\bibitem[{{Bunker} {et~al.}(2004){Bunker}, {Stanway}, {Ellis}, \&
  {McMahon}}]{bunker:04}
{Bunker}, A.~J., {Stanway}, E.~R., {Ellis}, R.~S., \& {McMahon}, R.~G. 2004,
  astro-ph/0403223

\bibitem[{{Bunker} {et~al.}(2003){Bunker}, {Stanway}, {Ellis}, {McMahon}, \&
  {McCarthy}}]{bunker:03}
{Bunker}, A.~J., {Stanway}, E.~R., {Ellis}, R.~S., {McMahon}, R.~G., \&
  {McCarthy}, P.~J. 2003, MNRAS, 342, L47

\bibitem[{{Cen}(2003)}]{cen:03}
{Cen}, R. 2003, ApJ, 591, 12

\bibitem[{{Cristiani} {et~al.}(2004)}]{cristiani:04}
{Cristiani}, S. {et~al.} 2004, ApJL, 600, L119

\bibitem[{{Dickinson} {et~al.}(2004)}]{dickinson:04}
{Dickinson}, M. {et~al.} 2004, ApJL, 600, L99

\bibitem[{{Dijkstra} {et~al.}(2004){Dijkstra}, {Haiman}, \&
  {Loeb}}]{dijkstra:04}
{Dijkstra}, M., {Haiman}, Z., \& {Loeb}, A. 2004, astro-ph/0403078

\bibitem[{{Djorgovski} {et~al.}(2001){Djorgovski}, {Castro}, {Stern}, \&
  {Mahabal}}]{djorgovski:01}
{Djorgovski}, S.~G., {Castro}, S., {Stern}, D., \& {Mahabal}, A.~A. 2001, ApJL,
  560, L5

\bibitem[{{Elvis} {et~al.}(2002){Elvis}, {Risaliti}, \& {Zamorani}}]{elvis:02}
{Elvis}, M., {Risaliti}, G., \& {Zamorani}, G. 2002, ApJL, 565, L75

\bibitem[{{Fan} {et~al.}(2003)}]{fan:03}
{Fan}, X. {et~al.} 2003, AJ, 125, 1649

\bibitem[{{Fan} {et~al.}(2004)}]{fan:04}
---. 2004, astro-ph/0405138

\bibitem[{{Floyd}(2004)}]{floyd:04}
{Floyd}, D.~J.~E. 2004, in Coevolution of Black Holes and Galaxies

\bibitem[{{Gammie} {et~al.}(2004){Gammie}, {Shapiro}, \&
  {McKinney}}]{gammie:04}
{Gammie}, C.~F., {Shapiro}, S.~L., \& {McKinney}, J.~C. 2004, ApJ, 602, 312

\bibitem[{{Giacconi} {et~al.}(2002)}]{giac:02}
{Giacconi}, R. {et~al.} 2002, ApJS, 139, 369

\bibitem[{{Giavalisco} {et~al.}(2004{\natexlab{a}})}]{giavalisco:04a}
{Giavalisco}, M. {et~al.} 2004{\natexlab{a}}, ApJL, 600, L93

\bibitem[{{Giavalisco} {et~al.}(2004{\natexlab{b}})}]{giavalisco:04b}
---. 2004{\natexlab{b}}, ApJL, 600, L103

\bibitem[{{Grimm} {et~al.}(2003){Grimm}, {Gilfanov}, \& {Sunyaev}}]{grimm:03}
{Grimm}, H.-J., {Gilfanov}, M., \& {Sunyaev}, R. 2003, MNRAS, 339, 793

\bibitem[{{Hansen} \& {Haiman}(2004)}]{hansen:04}
{Hansen}, S.~H. \& {Haiman}, Z. 2004, ApJ, 600, 26

\bibitem[{{Heger} {et~al.}(2003){Heger}, {Fryer}, {Woosley}, {Langer}, \&
  {Hartmann}}]{heger:03}
{Heger}, A., {Fryer}, C.~L., {Woosley}, S.~E., {Langer}, N., \& {Hartmann},
  D.~H. 2003, ApJ, 591, 288

\bibitem[{{Hui} \& {Haiman}(2003)}]{hui:03}
{Hui}, L. \& {Haiman}, Z. 2003, ApJ, 596, 9

\bibitem[{{Immler} {et~al.}(2004)}]{immler:04}
{Immler}, S. {et~al.} 2004, AJ, in prep.

\bibitem[{{Kennicutt}(1998)}]{kenn:98}
{Kennicutt}, R.~C. 1998, ARA\&A, 36, 189

\bibitem[{{Kogut} {et~al.}(2003)}]{kogut:03}
{Kogut}, A. {et~al.} 2003, ApJS, 148, 161

\bibitem[{{Loeb} \& {Barkana}(2001)}]{loeb:01}
{Loeb}, A. \& {Barkana}, R. 2001, ARA\&A, 39, 19

\bibitem[{{Madau} {et~al.}(1996){Madau}, {Ferguson}, {Dickinson}, {Giavalisco},
  {Steidel}, \& {Fruchter}}]{madau:96}
{Madau}, P., {Ferguson}, H.~C., {Dickinson}, M.~E., {Giavalisco}, M.,
  {Steidel}, C.~C., \& {Fruchter}, A. 1996, MNRAS, 283, 1388

\bibitem[{{Madau} {et~al.}(1999){Madau}, {Haardt}, \& {Rees}}]{madau:99}
{Madau}, P., {Haardt}, F., \& {Rees}, M.~J. 1999, ApJ, 514, 648

\bibitem[{{Madau} {et~al.}(2004){Madau}, {Rees}, {Volonteri}, {Haardt}, \&
  {Oh}}]{madau:04}
{Madau}, P., {Rees}, M.~J., {Volonteri}, M., {Haardt}, F., \& {Oh}, S.~P. 2004,
  ApJ, 604, 484

\bibitem[{{Magorrian} {et~al.}(1998)}]{magorrian:98}
{Magorrian}, J. {et~al.} 1998, AJ, 115, 2285

\bibitem[{{Malhotra} {et~al.}(2004)}]{malhotra:04}
{Malhotra}, S. {et~al.} 2004, ApJ, in prep.

\bibitem[{{Nandra} {et~al.}(2002)}]{nandra:02}
{Nandra}, K. {et~al.} 2002, ApJ, 576, 625

\bibitem[{{Pirzkal} {et~al.}(2004)}]{pirzkal:04}
{Pirzkal}, N. {et~al.} 2004, astro-ph/0403458

\bibitem[{{Reddy} \& {Steidel}(2004)}]{reddy:04}
{Reddy}, N.~A. \& {Steidel}, C.~C. 2004, ApJL, 603, L13

\bibitem[{{Sazonov} {et~al.}(2004){Sazonov}, {Ostriker}, \&
  {Sunyaev}}]{sazonov:04}
{Sazonov}, S.~Y., {Ostriker}, J.~P., \& {Sunyaev}, R.~A. 2004, MNRAS, 347, 144

\bibitem[{{Sokasian} {et~al.}(2003){Sokasian}, {Abel}, {Hernquist}, \&
  {Springel}}]{sokasian:03}
{Sokasian}, A., {Abel}, T., {Hernquist}, L., \& {Springel}, V. 2003, MNRAS,
  344, 607

\bibitem[{{Sokasian} {et~al.}(2004){Sokasian}, {Yoshida}, {Abel}, {Hernquist},
  \& {Springel}}]{sokasian:04}
{Sokasian}, A., {Yoshida}, N., {Abel}, T., {Hernquist}, L., \& {Springel}, V.
  2004, MNRAS, 350, 47

\bibitem[{{Somerville} {et~al.}(2003){Somerville}, {Bullock}, \&
  {Livio}}]{rss:03b}
{Somerville}, R.~S., {Bullock}, J.~S., \& {Livio}, M. 2003, ApJ, 593, 616

\bibitem[{{Somerville} {et~al.}(2004){Somerville}, {Lee}, {Ferguson},
  {Gardner}, {Moustakas}, \& {Giavalisco}}]{rss:04b}
{Somerville}, R.~S., {Lee}, K., {Ferguson}, H.~C., {Gardner}, J.~P.,
  {Moustakas}, L.~A., \& {Giavalisco}, M. 2004, ApJL, 600, L171

\bibitem[{{Somerville} \& {Livio}(2003)}]{rss:03a}
{Somerville}, R.~S. \& {Livio}, M. 2003, ApJ, 593, 611

\bibitem[{{Spergel} {et~al.}(2003)}]{spergel:03}
{Spergel}, D.~N. {et~al.} 2003, ApJS, 148, 175

\bibitem[{{Stanway} {et~al.}(2003){Stanway}, {Bunker}, \&
  {McMahon}}]{stanway:03}
{Stanway}, E.~R., {Bunker}, A.~J., \& {McMahon}, R.~G. 2003, MNRAS, 342, 439

\bibitem[{{Stanway} {et~al.}(2004)}]{stanway:04}
{Stanway}, E.~R. {et~al.} 2004, ApJL, 604, L13

\bibitem[{{Stark} {et~al.}(1992){Stark}, {Gammie}, {Wilson}, {Bally}, {Linke},
  {Heiles}, \& {Hurwitz}}]{stark:92}
{Stark}, A.~A., {Gammie}, C.~F., {Wilson}, R.~W., {Bally}, J., {Linke}, R.~A.,
  {Heiles}, C., \& {Hurwitz}, M. 1992, ApJS, 79, 77

\bibitem[{{Steidel} {et~al.}(1996){Steidel}, {Giavalisco}, {Pettini},
  {Dickinson}, \& {Adelberger}}]{steidel:96}
{Steidel}, C.~C., {Giavalisco}, M., {Pettini}, M., {Dickinson}, M., \&
  {Adelberger}, K.~L. 1996, ApJ, 462, L17

\bibitem[{{Steidel} \& {Hamilton}(1992)}]{steidel:92}
{Steidel}, C.~C. \& {Hamilton}, D. 1992, AJ, 104, 941

\bibitem[{{Stiavelli} {et~al.}(2004{\natexlab{a}}){Stiavelli}, {Fall}, \&
  {Panagia}}]{stiavelli:04a}
{Stiavelli}, M., {Fall}, S.~M., \& {Panagia}, N. 2004{\natexlab{a}}, ApJ, 600,
  508

\bibitem[{{Stiavelli} {et~al.}(2004{\natexlab{b}}){Stiavelli}, {Fall}, \&
  {Panagia}}]{stiavelli:04b}
---. 2004{\natexlab{b}}, astro-ph/0405219

\bibitem[{{Wyithe} \& {Loeb}(2003)}]{wyithe:03b}
{Wyithe}, J.~S.~B. \& {Loeb}, A. 2003, ApJ, 595, 614

\bibitem[{{Yan} \& {Windhorst}(2004)}]{yan:04}
{Yan}, H. \& {Windhorst}, R.~A. 2004, ApJL, 600, L1

\bibitem[{{Yu} \& {Tremaine}(2002)}]{yu:02}
{Yu}, Q. \& {Tremaine}, S. 2002, MNRAS, 335, 965

\end{thebibliography}

\end{document}